\pdfoutput=1

\documentclass[useAMS,usenatbib]{mn2e}

\usepackage{graphicx}

\title[Formation of a rotating jet]
{Formation of a rotating jet during the filament eruption on 10-11
April 2013}
\author[B. Filippov et al.]{B. Filippov,$^1$ \thanks{E-mail:
bfilip@izmiran.ru} A. K. Srivastava,$^2$ B. N. Dwivedi,$^2$ S.
Masson,$^3$ G. Aulanier,$^3$ \newauthor N. C. Joshi,$^4$ and W.
Uddin$^5$ \\ $^{1}$Pushkov Institute of Terrestrial Magnetism,
Ionosphere and Radio Wave Propagation of the Russian Academy of
Sciences (IZMIRAN), \\ Troitsk, Moscow 142190, Russia \\
$^{2}$Department of Physics, Indian Institute of Technology
(Banaras Hindu University), Varanasi, 221005, India \\
$^{3}$LESIA, Observatoire de Paris, CNRS, UPMC, Univ. Paris
Diderot, 5 place Jules Janssen, F-92190 Meudon, France \\
$^{4}$School of Space Research, Kyung Hee University, Yongin,
Gyeonggi-Do, 446-701, Korea \\ $^{5}$Aryabhatta Research Institute
of Observational Sciences (ARIES), Manora Peak, Nainital-263 009,
India}

\begin{document}

\date{Accepted 0000 December 15. Received 0000 December 14; in original form 0000 October 11}

\pagerange{\pageref{firstpage}--\pageref{lastpage}} \pubyear{2002}

\maketitle

\label{firstpage}

\begin{abstract}
We analyze multi-wavelength and multi-viewpoint observations of a
helically twisted plasma jet formed during a confined filament
eruption on 10-11 April 2013. Given a rather large scale event
with its high spatial and temporal resolution observations, it
allows us to clearly understand some new physical details about
the formation and triggering mechanism of twisting jet. We
identify a pre-existing flux rope associated with a sinistral
filament, which was observed several days before the event. The
confined eruption of the filament within a null point topology,
also known as an Eiffel tower (or inverted-Y) magnetic field
configuration results in the formation of a twisted jet after the
magnetic reconnection near a null point. The sign of helicity in
the jet is found to be the same as that of the sign of helicity in
the filament. Untwisting motion of the reconnected magnetic field
lines gives rise to the accelerating plasma along the jet axis.
The event clearly shows the twist injection from the pre-eruptive
magnetic field to the jet.
\end{abstract}


\section{Introduction}

Among the different types of plasma motions observed during solar
eruptive events in the corona and chromosphere, there are
apparently linear collimated plasma flows, which are presumably
guided by magnetic fields, usually referred to as jets. A wide
variety of jet-like structures are observed in the solar
atmosphere. They can be formed both from relatively cool plasma
such as spicules, spikes, macro-spicules and surges and from hot
plasma as seen in X-rays, white-light and EUV. Numerous
small-scale jets are observed by {\it Hinode}/X-ray telescope.

While the general shape of jets is a nearly straight or slightly
curved linear structure, high resolution and spectral observations
reveal complicated motions within many jets. Rotation about the
jet axis (spinning) and helical features were noticed long ago in
spicules \citep{b13,b49,b66}, surges \citep{b71,b32,b7}, and jets
\citep{b61,b47a,b36,b37,b55,b10,b53}. Such plasma flow behavior
can be expected in pre-existing twisted coronal magnetic flux
tubes or due to changes in the magnetic field during jet progress.
In the interpretation of low spatio-temporal resolution
observations, it is not easy to make right choice between these
two possibilities.

\begin{figure*}
\includegraphics[width=167mm]{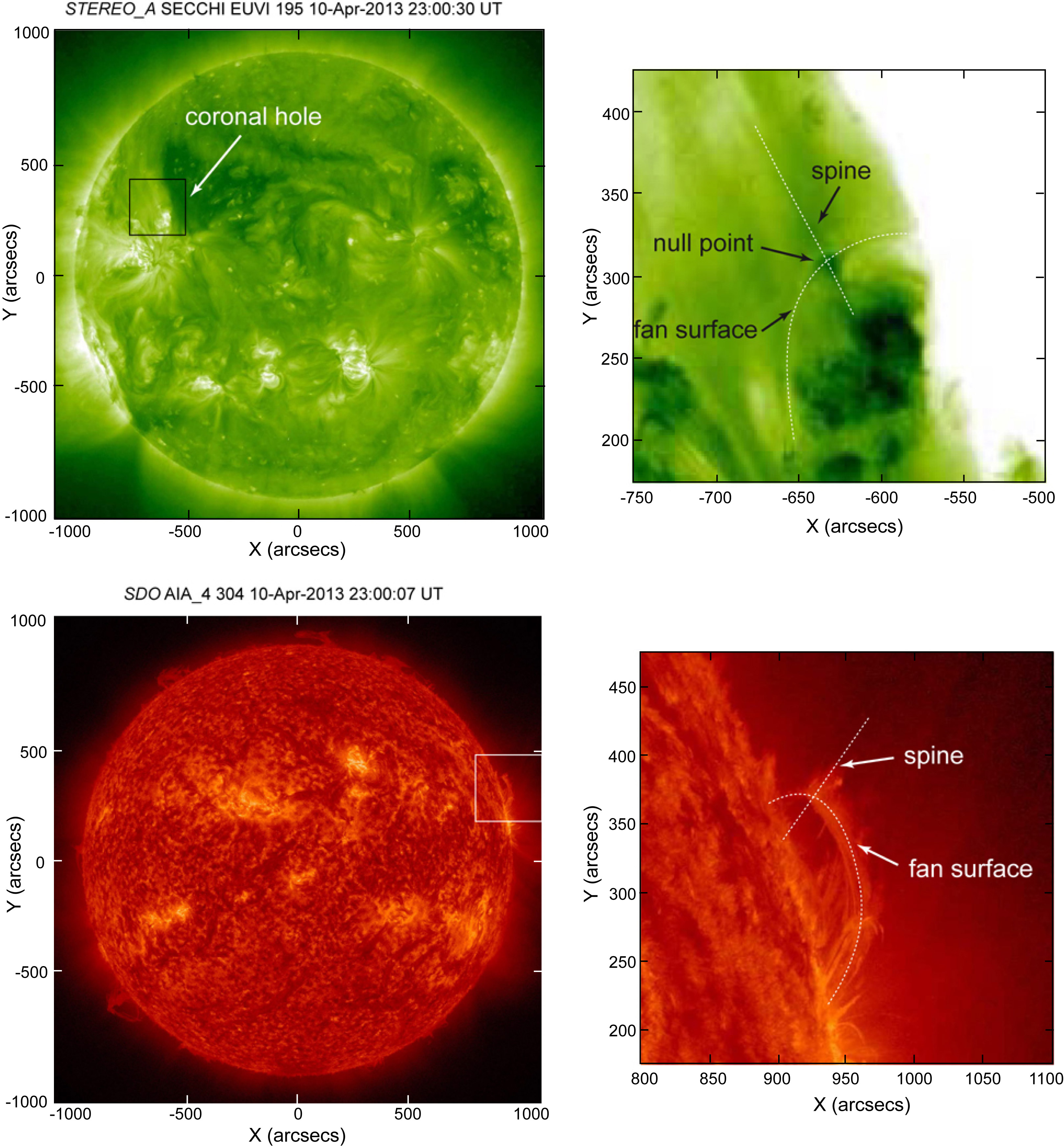}
\caption{Full disk {\it STEREO}/SECCHI/EUVI 195 \AA \ (top left),
{\it SDO}/AIA 304 \AA \ (bottom left) images and zoomed areas
(right panels) within boxes shown in left panels. }
\end{figure*}

\begin{figure*}
\includegraphics[width=167mm]{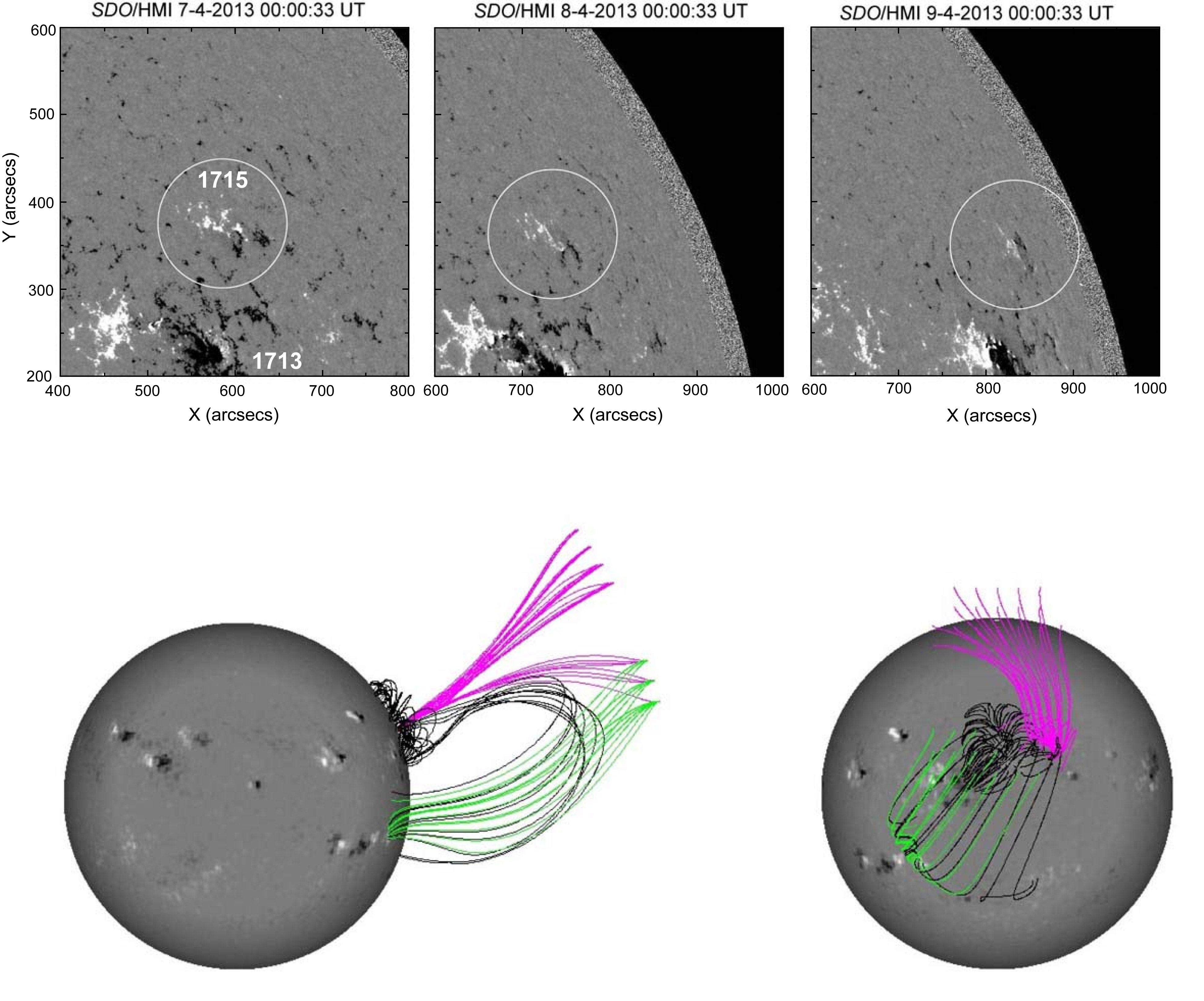}
\caption{Upper panel: {\it SDO}/HMI fragments of line-of-sight
magnetograms on 2013 April 7, 8 and 9. White circles show the
location of NOAA 1715 active region. Bottom panel: PFSS magnetic
field extrapolation showing a side view (left image) and a top
view (right image) of the coronal magnetic field structure over
NOAA 1715 active region. Magenta lines  (grey in B\&W images)
represent open coronal-hole field lines. Green lines  (light
grey in B\&W images) represent open field lines emanating from
the southern hemisphere.}
\end{figure*}

\begin{figure*}
\includegraphics[width=150mm]{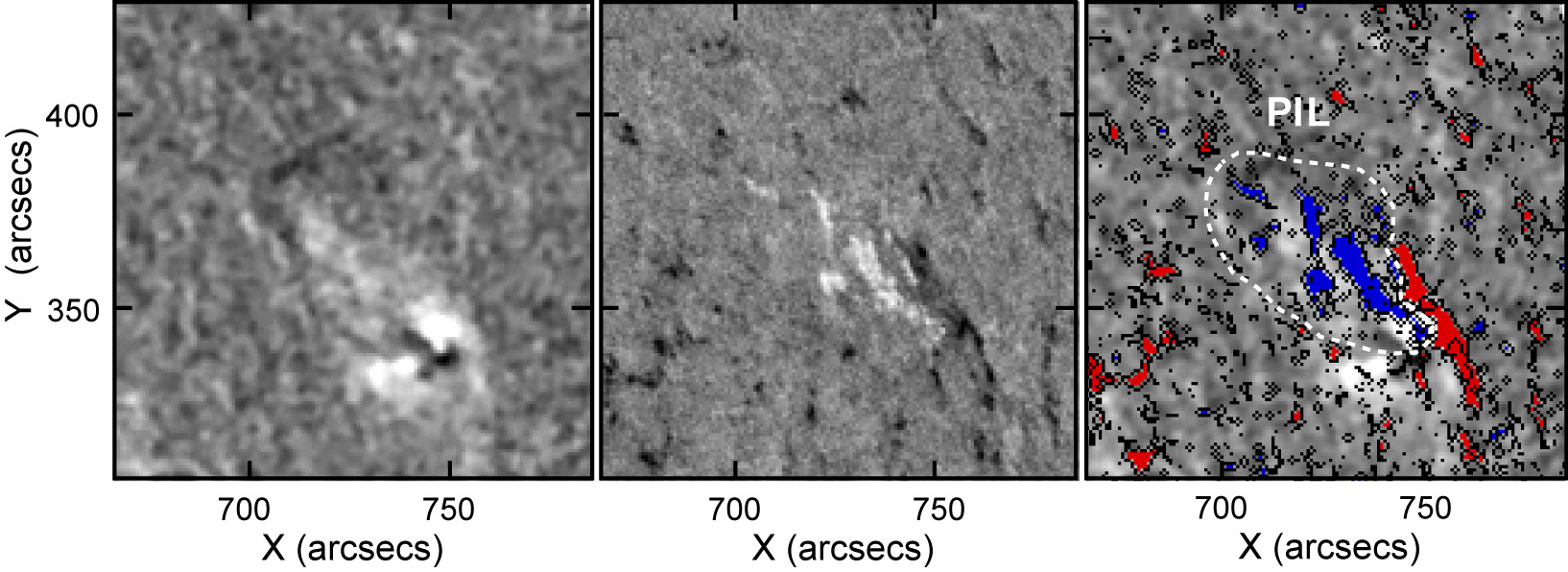}
\caption{A fragment of the H$\alpha$  filtergram taken at 09:52 UT
on April 08 at the Kanzelhoehe Solar Observatory (left panel),
{\it SDO}/HMI magnetogram of the same region (central panel),
H$\alpha$ filtergram with overlaid +50 gauss (G) magnetic field
contours from the {\it SDO}/HMI magnetogram (right panel). Red
areas  (light grey in the B\&W image) represent negative
polarity, while blue areas  (dark grey in the B\&W image)
represent positive polarity.  The white dashed line shows the
polarity inversion line (PIL).}
\end{figure*}

Nevertheless, the major characteristic of jets is collimated
field-aligned plasma motion. Plasma can be accelerated by the gas
pressure gradient force, which is enhanced by sudden heating in
the upper chromosphere \citep{b64,b67}, by the magnetic pressure
force in a relaxing magnetic twist \citep{b58}, or by the magnetic
tension force in the reconnection process \citep{b60,b57}.
Different models of jet formation based on reconnection have been
proposed in two-dimensional (2D) and three-dimensional (3D)
geometry \citep{b73,b25,b46}. Since the field line reconnection is
most favorable near magnetic null points, magnetic configurations
called as anemone-like, Eiffel tower, inverted-Y-shaped ones
containing null points are often considered as source regions for
jets. Reconnection is believed to be driven by emergence of a new
magnetic flux from below the photosphere. The emergence time-scale
is very slow compared with the time-scale of a jet phenomenon.
Therefore reconnection should be forbidden during the initial
phase of the flux emergence by low plasma resistivity \citep{b59}
or high symmetry \citep{b45}. In this phase, a large amount of
free magnetic energy is built up in the corona due to the
formation of current sheets. The reconnection is switched on by
anomalous resistivity or breaking the symmetry, which leads to an
explosive release of the free energy.

Some new jet observations are inconsistent in several features
with a `standard' scenario of jet formation when a bipolar
magnetic structure emerges into a pre-existing open magnetic field
and reconnects to form hot and fast outflows that are emitted from
the interface between the fields into contact \citep{b23}. They
were called `blowout jets' \ \citep{b40,b66,b37,b56}. Blowout jets
are broad, curtain-like structures in contrast to `standard' jets,
which are more elongated and contracting. In addition to the X-ray
emitting plasma, cooler plasma at chromospheric or transition
region temperatures erupts along these jets. The observed
structure and timing of these cooler ejecta suggest that this
plasma is erupted from low in the jet's base arch, in an ejective
eruption of the magnetic field in the core of the arch as in a
filament eruption and in a coronal mass ejection (CME). It is
believed that there is a twisted or a sheared arch in the source
region of a blowout jet, which often carries a small filament or
flux rope within it. When this structure becomes unstable and
erupts, it blows out the envelope field producing an untwisting
ejection of cool and hot material.

3D MHD simulations of the formation of jets \citep{b2} show that
external reconnection between emerging and ambient magnetic field
is important to turn on the standard jets, while the eruption of a
sheared field within the emerging region and internal reconnection
are key processes to form a twisted flux rope which is at the
origin of the blowout jet. Field lines are twisted along the jet
as a result of reconnection between the eruptive twisted field and
the pre-existing field, which consists of non-twisted oblique
field lines. Therefore, simulations show an untwisting plasma
motion during the emission of the blowout jet. In another
numerical experiment \citep{b41}, the launching of a hot and fast
coronal jet is followed by several violent eruptions. After the
standard jet phase with the reconnection process that takes place
at the interface between the colliding flux systems, a number of
violent eruptions of the magnetic field structure take place in
different volumes of the emerged plasma dome. Eruptions are
possibly caused by a kink or torus instability. During a flux-rope
expanding in height, its twist turns into writhe. \citet{b47}
studied the impact of the magnetic field inclination and
photospheric field distribution on the generation and properties
of straight and helical solar jets. They found that the 3D
magnetic null-point configuration is a very robust structure for
the energy storage and impulsive release characteristic of helical
jets. Reconnection occurring during the straight-jet phase
influences the triggering of the helical jet.

In this paper, we analyze new multi-wavelength and multi-viewpoint
observations of a helically twisted plasma jet formed during a
confined filament eruption on 10-11 April 2013. A rather large
scale of the event and high spatial and temporal resolution of the
{\it Solar Dynamic Observatory}/Atmospheric Imaging Assembly ({\it
SDO}/AIA; \citep{b34}) allow us to observe clearly some
comprehensive and new details of the jet formation. In Sect. 2, we
discuss general topology of the jet forming region. Dynamics and
fine structure of the observed jet is described and analyzed in
Sect. 3. Last section presents some discussions.

\section[]{General Topology of the Jet Source Region}

\begin{figure*}
\includegraphics[width=167mm]{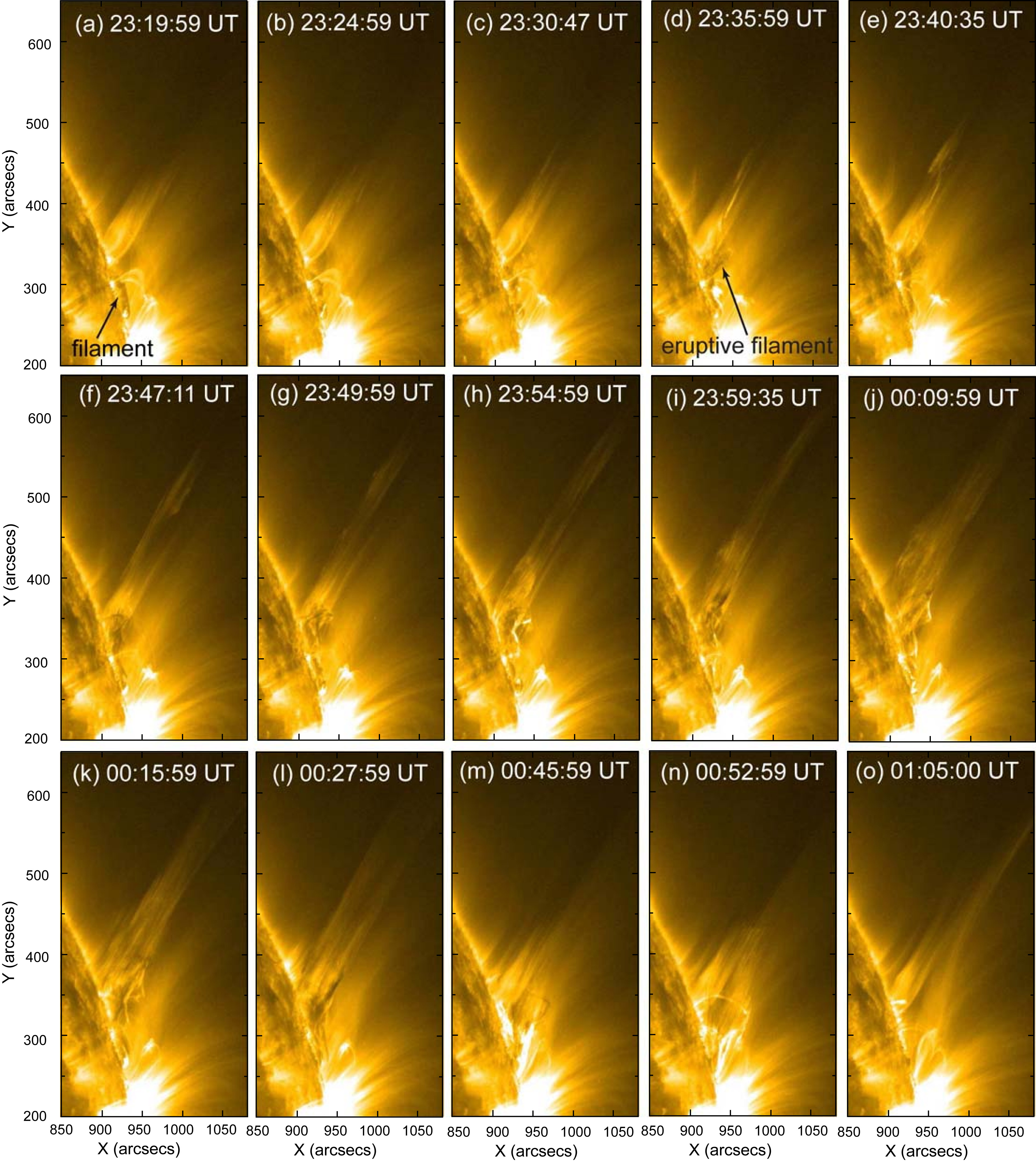}
\caption{Sequence of selected {\it SDO}/AIA 171~\AA \ images
showing a filament eruption and formation of a set of three
successive helically twisted jets. }
\end{figure*}

\begin{figure*}
\includegraphics[width=167mm]{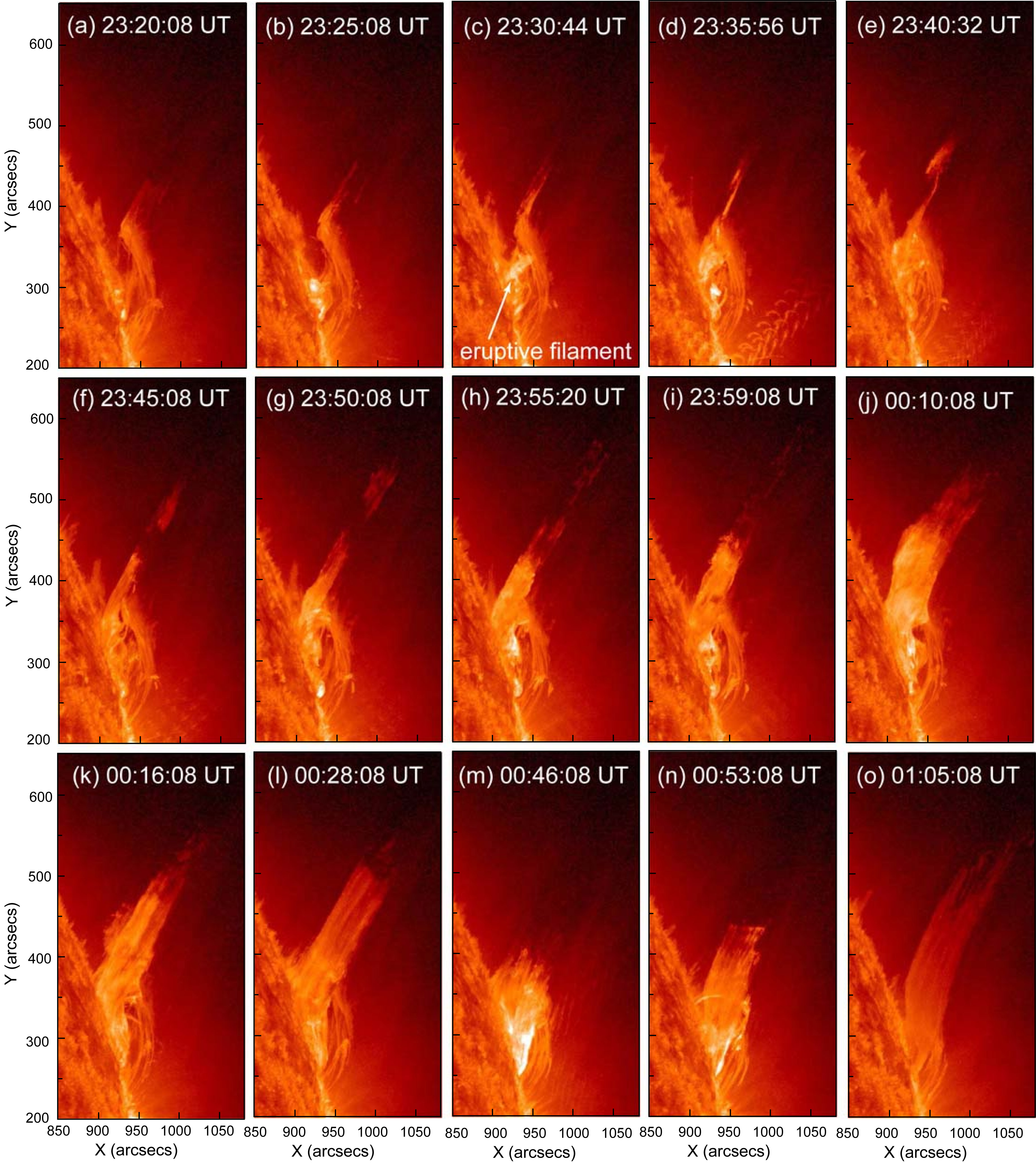}
\caption{Sequence of selected {\it SDO}/AIA 304~\AA \ images
showing a filament eruption and formation of a set of three
successive helically twisted jets.}
\end{figure*}

For this study, we have used multi-wavelength data from the AIA
and the Heliospheric and Magnetic Imager (HMI; \citep{b63})
onboard the {\it SDO}, the Sun Earth Connection Coronal and
Heliospheric Investigation (SECCHI) Extreme Ultraviolet Imager
(EUVI) \citep{b70,b24} on board the {\it Solar Terrestrial
Relations Observatory - Ahead} ({\it STEREO A}), the Large Angle
and Spectrometer Coronagraph (LASCO) \citep{b6}  on board the {\it
Solar and Heliospheric Observatory} ({\it SOHO}), and ground-based
full disk H$\alpha$ observations of the Kanzelhoehe Solar
Observatory. Observations from both the {\it STEREO-A}/SECCHI and
{\it SDO}/AIA provide a good opportunity to study a jet phenomenon
and associated dynamics on the disk as well as on the limb. The
separation angle of the {\it STEREO-A} with the Earth was
134$^\circ$ on 10-11 April 2013, accordingly a feature observed by
the {\it SDO} on the limb was located about 45$^\circ$ from the
central meridian for the {\it STEREO-A}.

Figure 1 shows the images of the Sun in the {\it STEREO}/EUVI 195
\AA \ (top panel) and {\it SDO}/AIA 304~\AA \ (bottom panel)
channels with boxes showing the region of jet-like ejection around
23:00 UT on 2013 April 10. Left panels show the full disk images
of the Sun, while the right panels show the zoomed images of the
area corresponding to the boxes in the left panels (the zoomed
{\it STEREO}/EUVI image is in negative colour). The jet was
associated with small National Oceanic and Atmospheric
Administration (NOAA) active region 1715 located to the north-west
from more developed NOAA active region 1713. NOAA 1715 active
region lies near the coronal hole identified as a dark area in the
{\it STEREO}/EUVI 195~\AA \ full disk image (top left panel of
Fig. 1). The shape of coronal loops appears as an Eiffel tower or
inverted-Y magnetic configuration of the jet source region. We can
distinguish a fan surface and a spine in both {\it STEREO} and
{\it SDO} images, which are shown by thin dashed white lines in
Fig. 1. A null point should exist at the intersection of the fan
surface and the spine. The coronal loops that outline the magnetic
configuration change insignificantly for many hours before and
after the jet event, which  indicates the stability of the
configuration. The studied jet event is only an episode in the
life of this region, which does not destroy its topology.

{\it SDO}/HMI photospheric magnetograms on 2013 April 7, 8 and 9
(Fig. 2, upper panel) show a patch of positive polarity surrounded
by a large area of negative polarity in this active region. The
coronal-hole base is also located in a negative polarity region.
The polarity inversion line associated with this region  (the
dashed line in Fig. 3) is nearly  circular and the connectivity
between the positive and the negative fluxes displays an
anemone-like shape. The potential-field source surface (PFSS)
magnetic field extrapolation \citep{b1,b52,b54} confirms the
presence of a null-point magnetic topology above NOAA 1715 active
region (Fig. 2, bottom panel).  High black lines connect NOAA
1715 active region with other regions and do not belong to the
fan-spine configuration. They are so high, therefore, they can be
considered as open field lines. The span of the fan surface in the
SDO images is about 0.15 solar radii, while the span of the closed
loop system in Fig. 2 is about 0.5 solar radii. In the bottom-left
panel of Fig. 1, it is seen that the spine is close to the
northern border of the structure. Therefore, the southern half of
the calculated loop system is only two times wider than the
southern half of the observed fan surface. We consider this as
rather good agreement taking into account rather low spatial
resolution of the PFSS magnetic field extrapolation and position
of the studied active region close to the limb (magnetic field in
the region cannot be measured on the day of jet observations). 

The closed flux is confined within a dome, the boundary surface of
which is the fan surface. Magenta lines  (grey in B\&W
images) represent the coronal-hole open field lines. According to
the directivity of the jet material, the outer spine is more
likely open and belongs to the coronal hole, but it could also be
alternatively closed among the large scale closed black field
lines south of the null-point \citep{b69}. Such null-point
topology agrees very well with the {\it STEREO} and {\it SDO}
observations highlighting an Eiffel tower configuration. Some
details of pre-eruptive configuration of the source region are not
very clear because of the position of the region being close to
the limb in both {\it SDO} and ground-based observations. They are
evidently clear only during the dynamical evolution of the event.

\begin{figure}
\includegraphics[width=84mm]{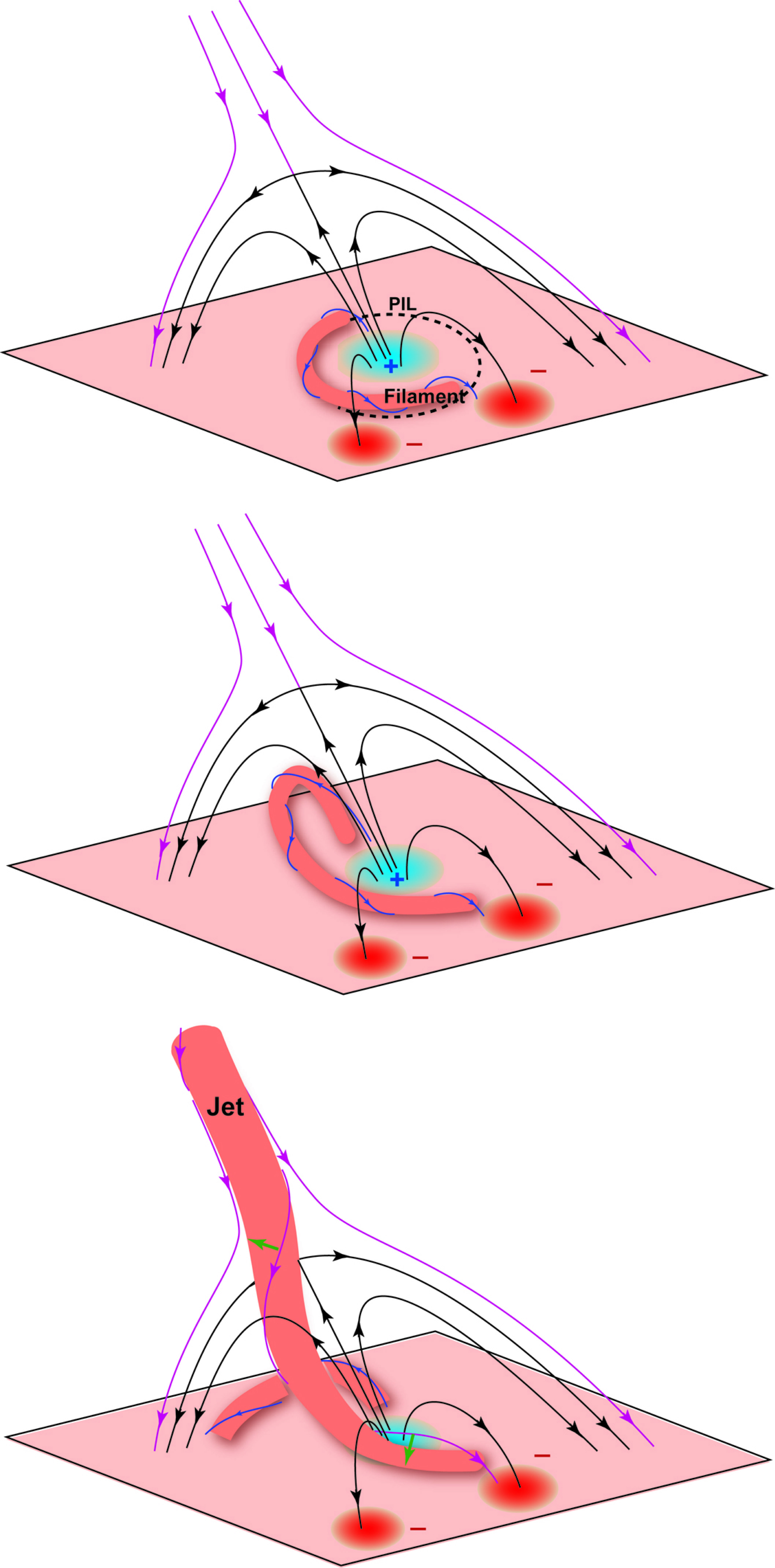}
\caption{Magnetic configuration of the jet source region derived
from {\it SDO} and {\it STEREO A} EUV images and photospheric
magnetic fields distribution (upper panel). Probable scenario of
the jet event (middle and lower panels). }
\end{figure}

A faint filament within NOAA 1715 active region can be recognized
in the H$\alpha$  filtergram taken at 09:52 UT on April 08 2013 at
the Kanzelhoehe Solar Observatory (Fig. 3). It is stretched along
the border of the patch of positive polarity within a large area
of negative polarity, which constitutes the base of a coronal
hole. There are stronger concentrations of negative polarity to
the south-west from the positive patch. However, it is not easy to
identify the polarity of each end of the filament where it is
connected. It is likely that the northern end of the filament is
connected to positive polarity, while the southern end is
connected to negative polarity. This conclusion is proven during
the eruption of the filament when the southern end remains
connected to the surface, while the northern part of the filament
forms a jet opening into the negative coronal hole. However, this
filament end connectivity means the sinistral chirality of the
filament, which is not typical for the northern hemisphere. The
hemispheric chirality rule is not strictly absolute\citep{b48} and
the conditions of its violation must be carefully analyzed. The
case of our filament seems to be exception from the general
hemispheric chirality rule.

The filament is clearly visible above the limb in {\it SDO}/AIA
171~\AA \ images on April 10, 2013 as a dark absorbing feature
with the top at a height of 13 Mm (Fig. 4 (a)). However, it is not
visible in {\it SDO}/AIA 304~\AA \ images before the first
filament activation at 23:20 UT when the first brightening appears
within the filament body.

Analysis of {\it SDO} and {\it STEREO A} EUV images and
photospheric magnetic field distribution shows that we have an
Eiffel tower or inverse Y magnetic configuration in the corona
(e.g. \citep{b57}) with a null point. The fan surface is outlined
by stable loops in {\it SDO}/AIA 304~\AA \ images (Fig. 1 and Fig.
5) with anchored southern ends and suspended northern ends just
near the probable null point position. The direction of the spine
slightly inclined to the north from the vertical becomes obvious
after the jet formation, although it can be vaguely recognized
from the direction of long straight loops visible in {\it SDO}/AIA
171~\AA \ images on the background (Fig. 4). From the 304~\AA \
image at 23:35:56 UT, we can estimate a height of the null point
above the limb and the size of the dome along the limb as 40~Mm
and 100~Mm respectively.

PFSS magnetic field extrapolation (Fig. 2) supports the schematic
magnetic configuration shown in Fig. 6. Open field lines belonging
to the coronal hole are shown with the magenta color  (grey in
B\&W images) in both figures, while closed field lines below the
fan surface are black. One of the helical field lines composing a
flux rope, which contains the filament, is shown with the blue
color before it becomes open after the jet formation and obtains
the magenta color.

\section[]{Dynamics and Fine Structure of Jets}

\begin{figure*}
\includegraphics[width=150mm]{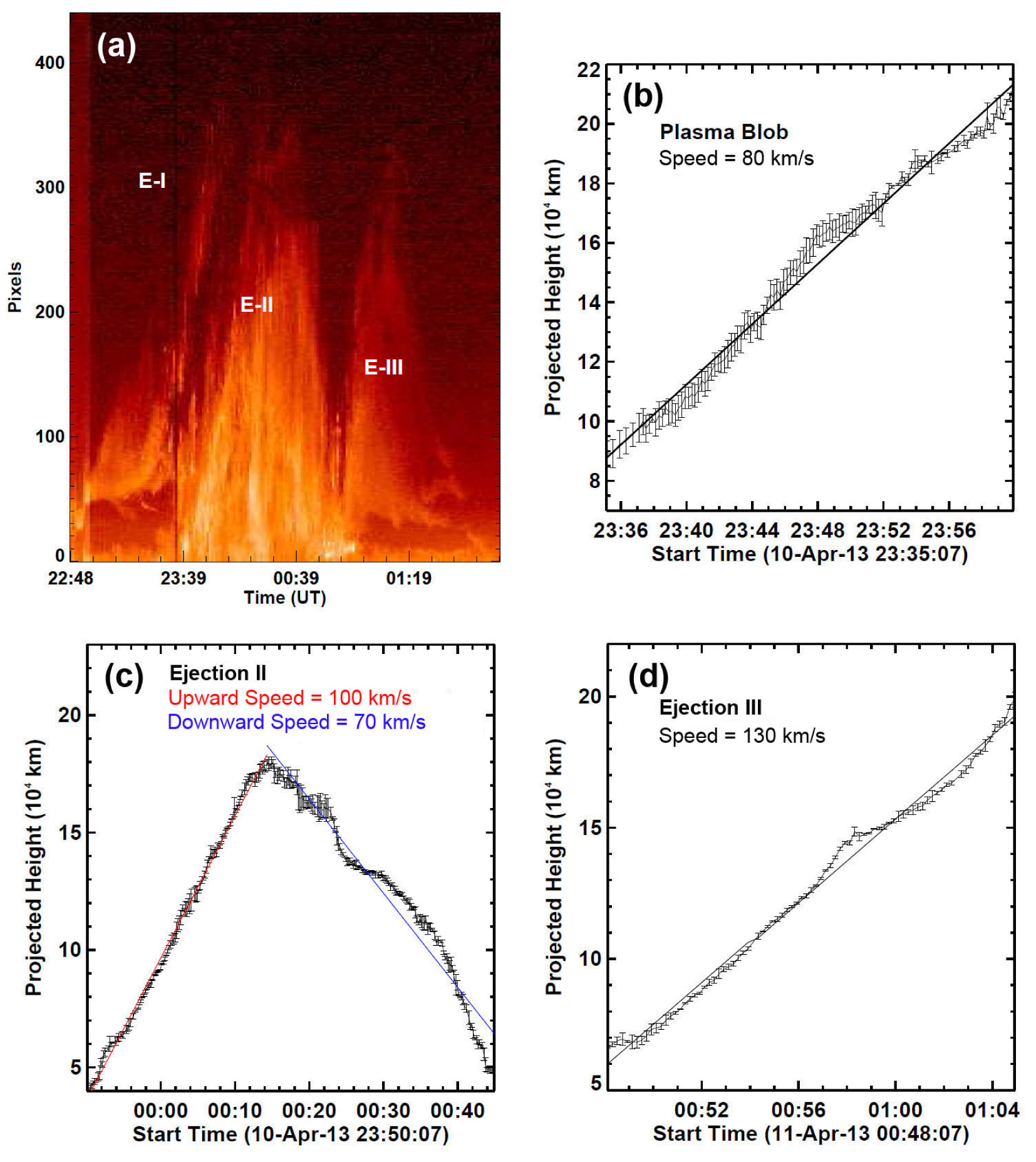}
\caption{Time-slice diagram from cuts along the spine showing
three ejections in the {\it SDO}/AIA 304~\AA \ channel (a),
kinematics of the three jets (b) -(c).}
\end{figure*}

\begin{figure*}
\includegraphics[width=120mm]{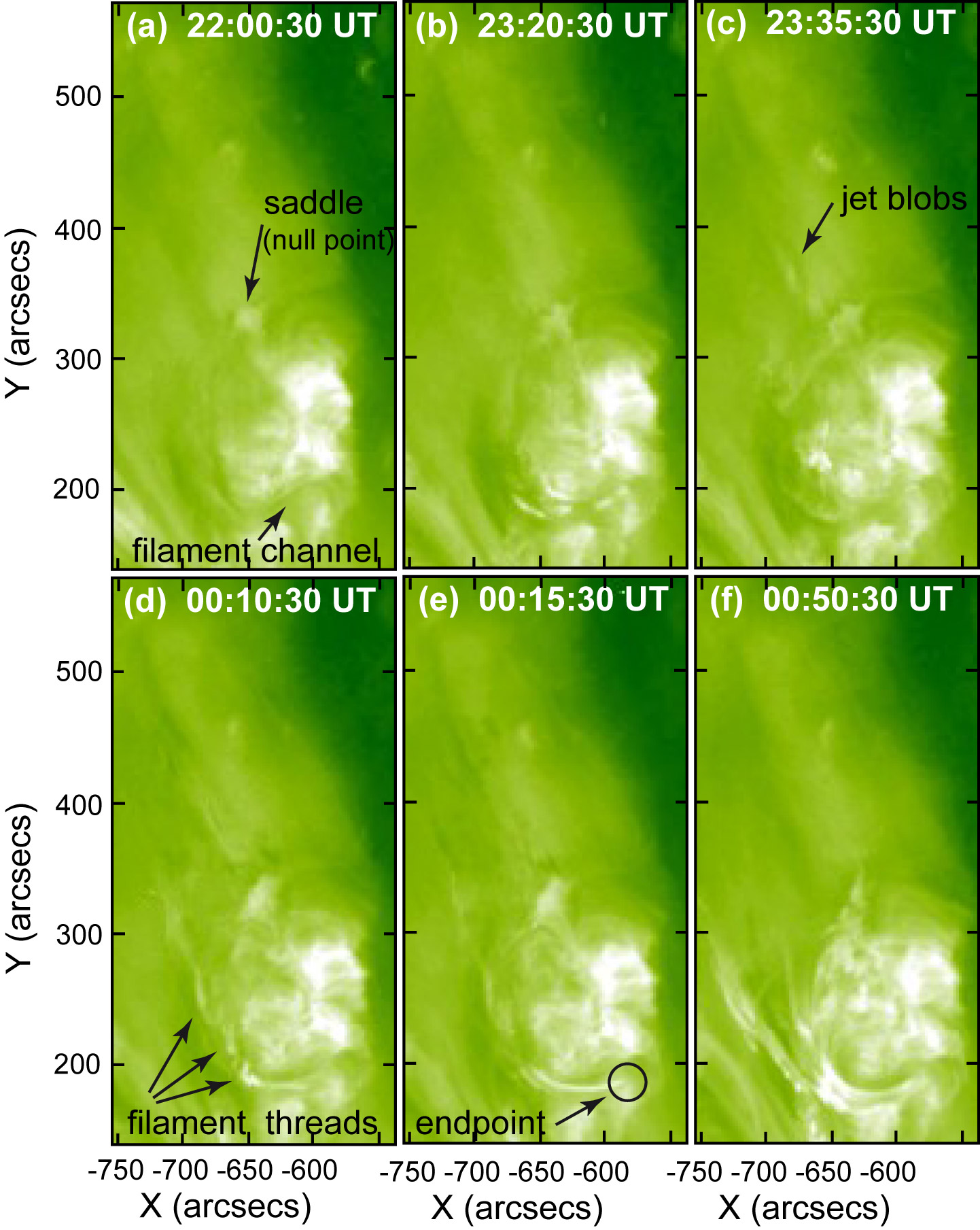}
\caption{Selected {\it STEREO}/EUVI 195~\AA \ images showing the
filament eruption and jet formation.}
\end{figure*}

\begin{figure}
\centering
\includegraphics[width=87mm]{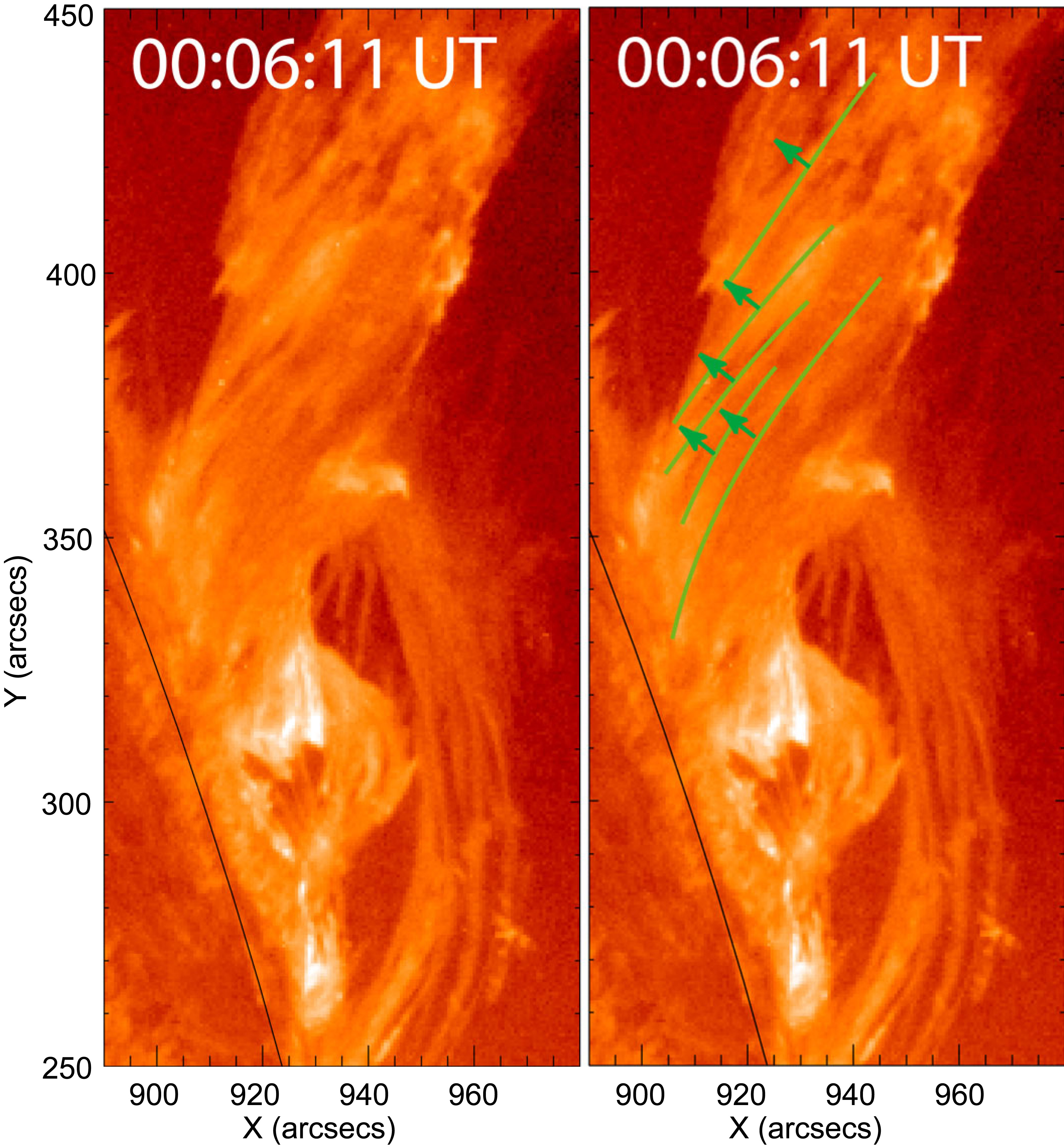}
\caption{{\it SDO}/AIA 304~\AA \ image of the jet at 00:06:11 UT.
The most conspicuous threads on the frontal side of the jet are
shown by green lines in the right panel. Their displacements
clearly seen in movie are in directions pointed by short green
arrows.}
\end{figure}

Figures 4 and 5 represent the temporal sequences of selected
images showing the overall dynamics of eruption and successive jet
formation in the {\it SDO}/AIA 171 and 304~\AA \ channels (see
also movie\_171 and movie\_304). The event starts from a
brightening at the southern end of the filament at 23:20 UT
followed by the rising of a bright loop over the middle part of
the filament. The loop does not rise over the fan surface but
seemingly it initiates the first blobs moving along the spine with
rotation around it. Dark filament material in {\it SDO}/AIA
171~\AA \ images moves to the north and rises there just below the
implied coronal null point. In few minutes, the filament
accelerates to a speed of the order of 100~km~s$^{-1}$, then
decelerates to a speed of about 40~km~s$^{-1}$, and reach the
height of 35~Mm, slightly below the fan surface, at 23:50 UT with
even lower speed. After this time, it is difficult to recognize
the top of the filament because it starts to transform into the
jet. Some parts of the filament plasma are heated and becomes
bright (emitting) in the {\it SDO}/AIA 171 and 304~\AA \ channels.
This mixture of cooler and hotter plasma starts rising along the
spine direction with the clockwise rotation (if viewed from top).

Three pulses of plasma acceleration along the spine can be
identified during the event.  We fix the position of the
leading edge of the jet and measure its coordinate along the axis
of the jet (Fig. 7). The errors (standard deviations) were
estimated by the three repeated measurements of the same jet's
leading edge or rotating bright point. The first flow was
launched by the rising bright loop at about 23:30 UT mostly as
rotating separate blobs bright both in the 171 and 304~\AA \
channel. An average speed of the blob motion along the spine has
been found to be about 80~km~s$^{-1}$ (Fig. 7(b)). No traces of
downward motion have been observed after this jet.  The second jet
is initiated by the dark filament eruption after 23:50 UT. It
contains bright as well as dark long threads rotating clockwise.
The dark threads have been observed not only in 171~\AA \ images
but also in 304~\AA \ images, which reveal the presence of cold
material. After reaching a height of 180~Mm around 00:14 UT on
2013 April 11, the most of the plasma material fall down. An
average speed of rising motion is about 100~km~s$^{-1}$, while a
speed of falling down plasma is at 70~km~s$^{-1}$ (Fig. 7(c)). The
third jet started after a new most spacious brightening near the
southern end of the filament at 00:35 UT. This jet is seen  most
elongated and show the greatest rising averaged speed of
130~km~s$^{-1}$ (Fig. 7(d)).

{\it STEREO}/EUVI observations show activation of the filament
within a filament channel (Fig. 8). Bright features appear in the
southern section of the filament and propagate along the filament
channel to the north. Saddle-like structure (Fig. 8(a)) projected
to the north of the active region marks the position of the null
point. The direction of the spine should coincide with the
direction of the ambient coronal magnetic field revealed by faint
coronal loops (north-north-east in this projection). The first and
second jets propagate along the spine with the saddle-like
structure at the base. The jets arise when bright features
spreading along the filament reach the vicinity of the saddle.
Bright blobs and thin dark threads can be easily recognized in
{\it STEREO}/EUVI 195~\AA \ images (Fig. 8(c)-(e)). The main body
of the third jet is rather far from the saddle structure. In this
case, the connection of the jet with the southern endpoint of the
filament is most obvious (Fig. 8(f)), although it can also be
recognized in the two other jets.

\begin{figure*}
\includegraphics[width=167mm]{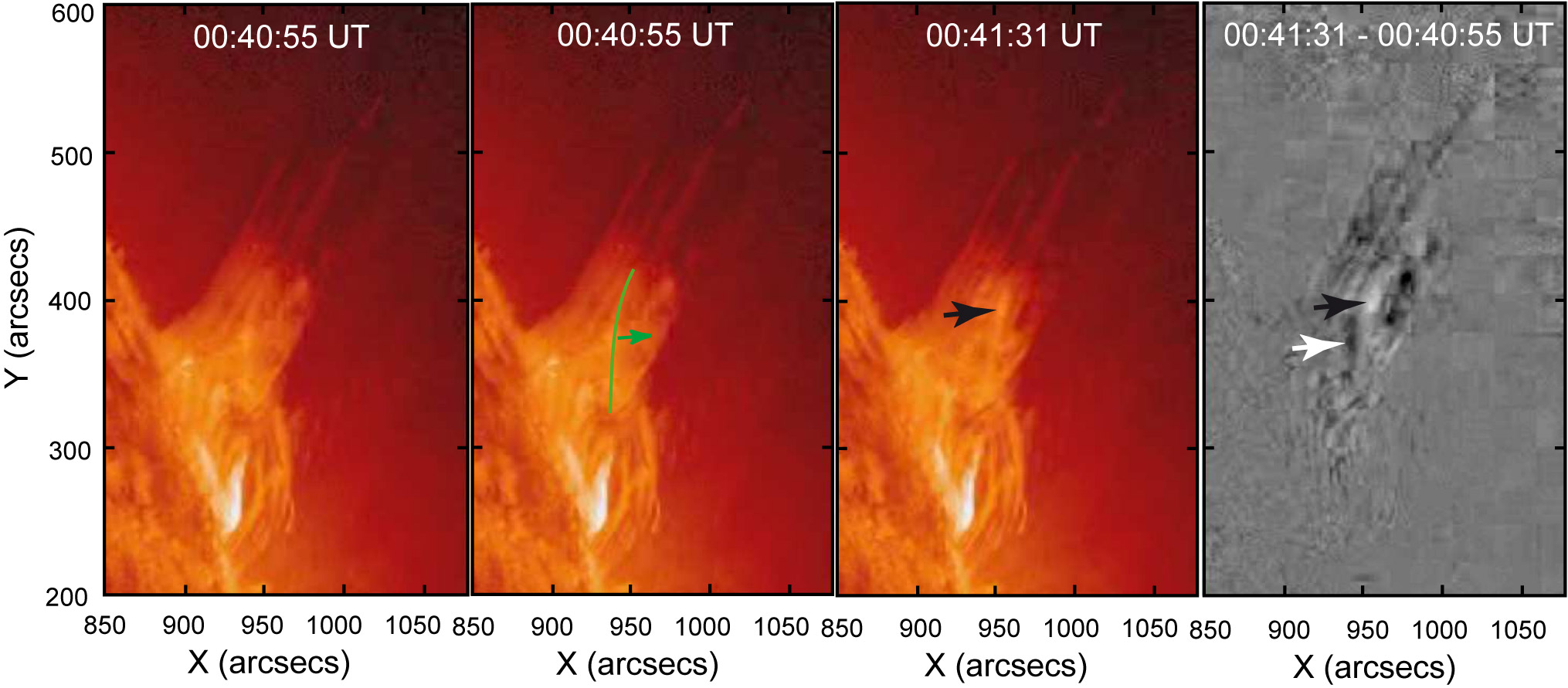}
\caption{{\it SDO}/AIA 304~\AA \ images of the jet at 00:40:55 UT
and 00:41:31 UT. The bright thread marked by green line in the
00:40:55 UT snapshot and pointed by the black arrow in the
00:41:31 UT snapshot is located on the far side of the jet and
moves from left to right as shown by the short green arrow.
Black/white features pointed to by white/black arrows in the
difference image in the right panel correspond to the previous and
subsequent positions of the thread.}
\end{figure*}

\begin{figure*}
\includegraphics[width=140mm]{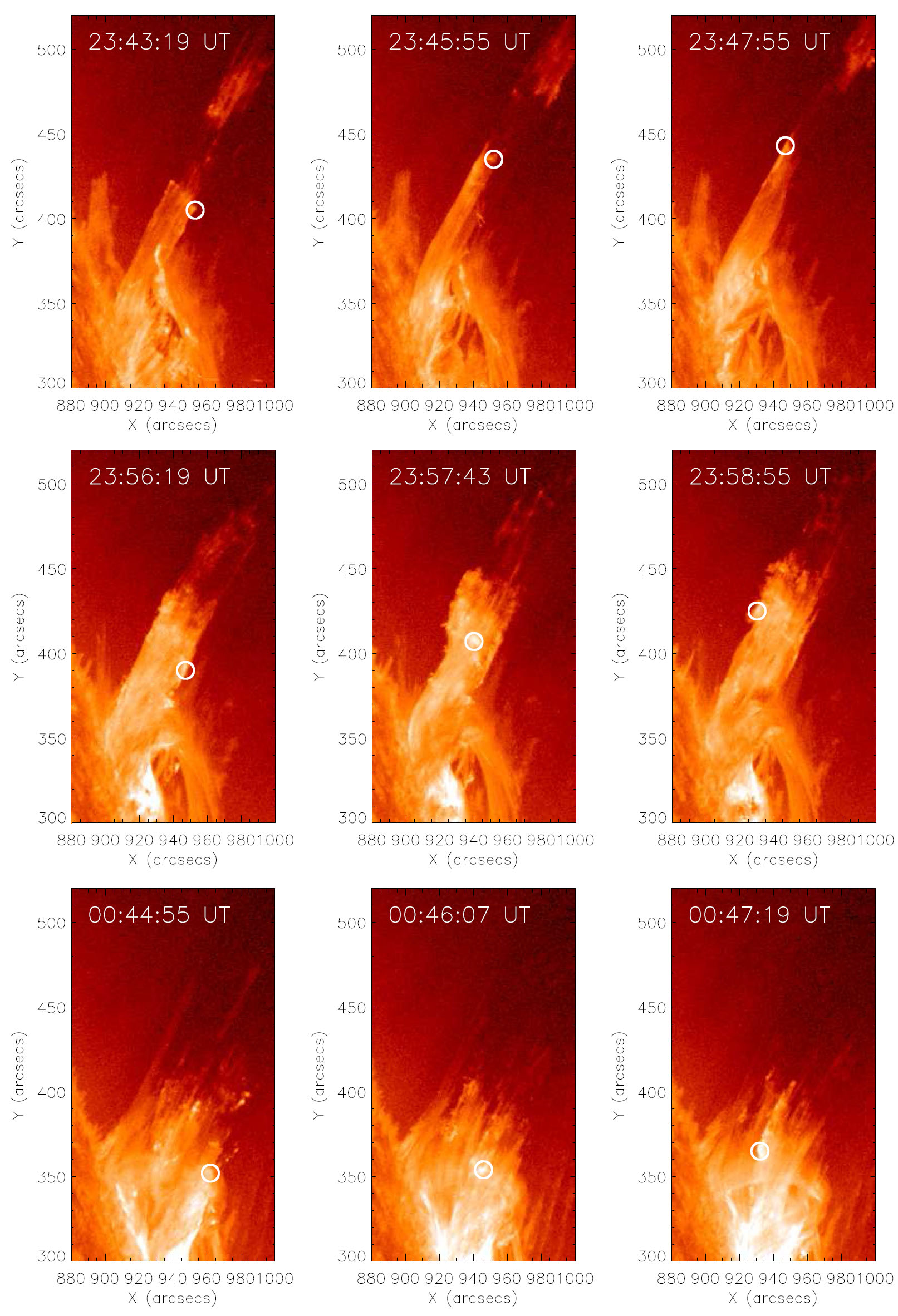}
\caption{{\it SDO}/AIA 304~\AA \ images showing the displacement
of tiny bright features (within white circles) moving around the
plasma column during the first (upper panel), second (middle
panel) and third (bottom panel) jet-like ejection. These images
were used to estimate the linear speed of rotation within jets.}
\end{figure*}

\begin{figure*}
\includegraphics[width=167mm]{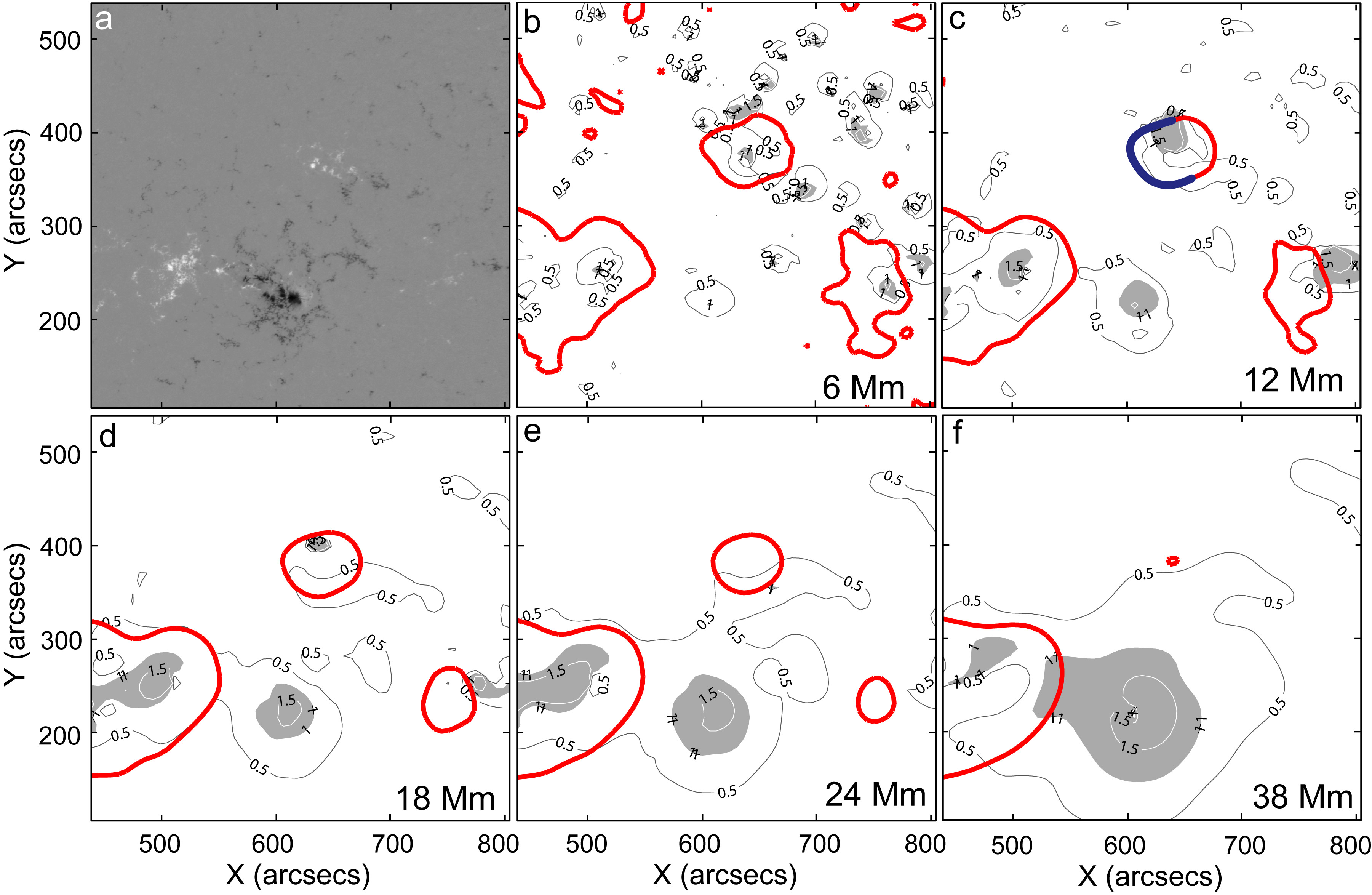}
\caption{(a) HMI magnetogram of active regions NOAA 1713 and NOAA
1715 taken on 2013 April 07 at 05:00 UT (Courtesy of NASA/{\it
SDO} and the HMI science team.) (b) - (f) Distributions of the
decay index and PILs (thick red lines) at different heights.
Shadowed areas show the regions where $n > 1$. The section of the
PIL occupied by the filament is marked with blue in (c) panel. }
\end{figure*}

The flux rope associated with the filament has the right-handed
screw (positive helicity) in accordance with the sinistral
chirality of the filament \citep{b48,b25a}. Bright filament
threads during its activation (Fig. 8 (d), (e)) deviate clockwise
from the filament axis. If they represent top sections of the
helical flux tubes, as expected for the activated filament, the
screw is really right-handed. Figures 9, 10 and movies show
clearly the right-handed screw in the fine structure of the jet
column. This indicates an unwinding of the pre-existed flux rope
associated with the filament. Such unwinding is expected for an
erupting flux-rope loop, since its length is increasing, while the
number of helical turns remains the same.

Rotational motion of plasma blobs could also be observed in a
twisted static magnetic field due to field aligned motion of the
blobs. We believe that in our case we observe the rotation of the
whole magnetic structure because long thin threads displace as a
whole from right to left on the frontal side of the jet and from
left to right on the far side of the jet as shown in Fig. 9 and
Fig. 10. Then, we measured the linear speed of rotation within
jets using the displacement of tiny bright features (Fig. 11) and
compared it with the averaged axial speed of plasma flows. The
linear speed of rotation within the first jet was about
100~km~s$^{-1}$, higher than speed along the spine
(85~km~s$^{-1}$). The same relationship was typical also for the
second jet: 180~km~s$^{-1}$ and 100~km~s$^{-1}$ and the third jet:
200~km~s$^{-1}$ and 130~km~s$^{-1}$. Taking into account the width
of the jet column of about 30 Mm, the angular speed of three jets
is 1.1 10$^{-3}$, 1.9 10$^{-3}$, and 2.1 10$^{-3}$ rad s$^{-1}$,
respectively. Since the linear speed of rotation significantly
exceeds the jet propagation speed, while the threads make acute
angles with the jet axis, magnetic field lines move (rotate) more
rapidly than plasma moves along them.

A narrow coronal mass ejection (CME) appeared in the north-west
sector of the {\it SOHO}/LASCO C2 field-of-view at 00:00 UT on 11
April. Obviously that with an initial speed of $\approx$
100~km~s$^{-1}$ this slow CME started long before the jet
ejection. However, its position angle was very close to the
position angle of the jets, thus CME's material masked a possible
continuation of the jets into the upper corona.

\section[]{Discussion}

In the jet event on 10-11 April 2013, we observe the rising motion
of the top of the filament (presumably showing the flux-rope axis)
only up to the fan surface. Then plasma moves along the general
direction of the surrounding coronal magnetic field forming the
jet. We can assume that the flux-rope magnetic field reconnects
with the open coronal field somewhere near the coronal null point
(Fig. 6). The twisted flux rope was cut in the middle part or
closer to the northern end, and two halves find new connections.
The southern end of the flux rope is still anchored in negative
polarity to the south from the positive patch, while the other
ends of these flux tubes are open to the outer corona. The
southern part of the flux rope is retained at an approximately
initial height by the stronger field related to the photospheric
negative polarity concentrations. The northern half of the flux
rope should find connection with negative polarity somewhere below
the fan surface.

We have no information about evolution of photospheric magnetic
fields at the base of jets just before and during their formation
because the region was on the limb for magnetographic
observations. Thus, we cannot relate the jet formation with
emergence of new magnetic flux, as it is assumed in the majority
of jet models (as well as cannot deny such relation). On the other
hand, we have evidence of the presence of a flux rope in the
active region before the jet event and evidence of flux-rope
instability. The flux rope is disclosed by the filament. There is
little doubt that filaments are enclosed within flux ropes. We can
estimate the stability of flux rope equilibrium in the coronal
magnetic field knowing its height above the photosphere and
calculating the so-called decay index $n$
\citep{b4,b17,b18,b29,b15,b74}. The stable equilibrium is possible
only if the background field decreases with height not too fast,
or the decay index of the ambient magnetic field does not exceed a
critical value $n_c$. \citet{b15} have shown that the critical
decay index $n_c$ has values in the range 1.1 - 1.3, if a flux
rope expands during an upward perturbation, and in the range 1.2 -
1.5, if a flux rope does not expand. Horizontal equilibrium of a
coronal electric current (a flux rope) is possible only at a
polarity inversion line (PIL) in the corona.

Figure 12(a) represents a fragment of the magnetogram taken by the
{\it SDO}/HMI on 2013 April 7 at 05:00 UT, which was used as a
boundary condition for the potential magnetic field calculations.
 We solve the Neuman external boundary-value problem
numerically using the Green function method \citep{b17,b18,b16}.
Since we need the magnetic field distribution in the corona at
heights of prominences, which is much less than a solar radius, we
use only a restricted area of a photospheric magnetogram as the
boundary of the Cartesian domain and neglect its sphericity
considering as a part of a flat surface. We cut out a rectangular
area around active region NOAA 1715 from the full disk magnetogram
including also neighbor active region NOAA 1713 as a strong source
of the magnetic field in the corona near the filament. The size of
the box as seen in Fig. 12 is 600$^{\prime\prime}$ $\times$
577$^{\prime\prime}$. Since the pixel size in HMI magnetograms is
about 0.5$^{\prime\prime}$, the size of the data array is 1200
$\times$ 1154. Summation over all these points in every point of
the map would need rather great computational time. However we
found that for the magnetic field parameters at heights above 6 Mm
the results were not changed noticeably after several binning, up
to the pixel size of 8$^{\prime\prime}$.

Then we calculate the distribution of the decay index $n$ =
-dln$B$/dln$h$ in horizontal surfaces $h$ = const at different
heights $h$. In Fig. 12(b) - (f), the thin lines show isocontours
of $n = 0.5, 1, 1.5$, while the thick red lines indicate the
positions of PILs at respective heights. Areas where $ n > 1$ are
shadowed. Blue line in Fig. 12(c) shows the section of the PIL
occupied by the filament as it is seen in H$\alpha$ line (Fig 3).
Within a height interval from 10 to 20~Mm, there is a small volume
of high values of the decay index ($n > 1.5$) near the northern
end of the filament. Therefore, vertical equilibrium is not stable
for the filament with the height of 13~Mm (Fig. 4(a)). It should
start to erupt. However, the stable equilibrium is possible again
at higher heights ($\sim$~20~Mm). We can expect a confined
eruption of the filament (the flux rope) beginning in its northern
section, as it is observed. The southern section of the flux rope
is more stable because the decay index is small there ($n < 1$).

During such slow eruption, the flux rope stays in
quasi-equilibrium. For horizontal equilibrium, it should be
located near the magnetic neutral surface or at a PIL at any
height. The neutral surface in NOAA 1715 has a dome-like shape as
the fan surface with a smaller horizontal size than the fan
surface has at the same height. The neutral surface is situated
below the fan surface. They touch each other at the top, in the
null point. Our potential-field calculations show the neutral
surface summit at a height of 40~Mm (Fig. 12(f)), while the height
of the fan surface top obtained from {\it SDO}/AIA 304~\AA \
images is also about 40~Mm. Hence, the upper part of the slowly
erupting flux rope comes to the vicinity of the null point, where
there are favorable conditions for magnetic field line
reconnection. Three consecutive jet ejections possibly correspond
to three consecutive reconnections of separate flux tubes
constituting the flux rope with open field lines above the fan
surface.

We do not observe any traces of energy release near the null point
during the jet formation. Bright features appear in both {\it
STEREO} and {\it SDO} images within the filament body starting
from the vicinity of the southern endpoint. The only observational
manifestation of reconnection is a change in the field lines
topology. It seems that magnetic untwisting \citep{b58,b45,b46} is
the driving mechanism of plasma acceleration along the spine. One
of the reasons is that the linear speed of rotation of jet
features is higher than the axial speed of the jet. The sign of
helicity observed in the jets corresponds to the sing of helicity
of the filament, which is derived from its chirality. In this
event, we have a clear example of helicity transport from a closed
flux to an open flux.

Interaction of different magnetic fluxes is the necessary
condition of reconnection. Usually, emergence of new flux is
considered as a driver of reconnection and jet formation, although
in some jet events there is no evidence of photospheric magnetic
field changes during ejections \citep{b9}, or flux decrease is
observed \citep{b11}. Only small changes in flux distribution
showing both flux increase and decrease or slow motion of flux
concentrations were found in association with jets
\citep{b37,b53}. On the other hand, rapid changes in the flux
distribution in the corona can be caused by instabilities of
pre-existing coronal flux ropes. The presence of flux ropes in jet
source regions and their eruption during jet formation is supposed
in many blowout jet events \citep{b19,b40,b37,b55,b53} and used as
a jet driver in many models \citep{b2,b33}. However, observations
of the pre-existing flux ropes (as filaments) are few
\citep{b56,b20}. In the jet event on 10-11 April 2013, we can
identify the sinistral filament, which was observed several days
before the event. The confined eruption of the filament within
Eiffel tower or inverted-Y magnetic configuration led to formation
of the twisted jet after reconnection at the null point between
the flux rope and the external open field. The sign of helicity in
the jet is the same as the sign of helicity in the filament.
Untwisting motion of reconnected field lines accelerates plasma
along the jet axis. The event clearly shows the twist injection
from the closed pre-eruptive magnetic field of the filament to the
open magnetic field of the jet.

\section*{Acknowledgments}

The authors acknowledge the Kanzelhoehe Solar Observatory, {\it
STEREO}, {\it SOHO}, and {\it SDO} teams for the high-quality data
supplied. This work was supported in part by the Russian
Foundation for Basic Research (grant 14-02-92690) and in part by
the Department of Science and Technology, Ministry of Science and
Technology of India (grant INT/RFBR/P-165). BPF also acknowledges
visit to Department of Physics, IIT (BHU) in November 2014 where
the parts of this present work is carried out.

\bsp

\label{lastpage}


\begin{thebibliography}{99}
\bibitem[\protect\citeauthoryear{Altschuler \& Newkirk}{1969}]{b1}
Altschuler, M.D., Newkirk, G., 1969, Sol. Phys. 9, 131
\bibitem[\protect\citeauthoryear{Archontis \& Hood} {2013}]{b2}
Archontis, V., Hood, A.W., 2013, ApJ, 769, L21
\bibitem[\protect\citeauthoryear{Bateman}{1978}]{b4}
Bateman G., 1978, MHD Instabilities, Massachusetts Institute of
Technology, Cambridge, MA
\bibitem[\protect\citeauthoryear{Brueckner et al.}{1995}]{b6}
Brueckner G.E. et al., 1995, Sol. Phys., 162, 357
\bibitem[\protect\citeauthoryear{Canfield et al.}{1996}]{b7}
Canfield, R.C., Reardon, K.P., Leka, K.D., Shibata, K., Yokoyama,
T., Shimojo, M., 1996, ApJ, 464, 1016
\bibitem[\protect\citeauthoryear{Chandrashekhar et al.}{2014}]{b9}
Chandrashekhar, K., Morton, R.J., Banerjee, D., Gupta, G.R., 2014,
A\&A, 562, A98
\bibitem[\protect\citeauthoryear{ Chen, Zhang \& Ma}{ Chen et al.}{2012}]{b10}
Chen, H.-D., Zhang, J., Ma, S.-L., 2012, RA\&A, 12, 573
\bibitem[\protect\citeauthoryear{Chifor et al.}{2008}]{b11}
Chifor, C., et al., 2008, A\&A, 491, 279
\bibitem[\protect\citeauthoryear{Cook et al.}{1984}]{b13}
Cook, J.W., Brueckner, G.E., Bartoe, J.-D.F., Socker, D.G., 1984,
Adv. Space Res., 4, 59
\bibitem[\protect\citeauthoryear{ D$\acute{e}$moulin \& Aulanier}{2010}]{b15}
D$\acute{e}$moulin P., Aulanier G., 2010, ApJ, 718, 1388
\bibitem[\protect\citeauthoryear{Filippov}{2013}]{b16}
Filippov B., 2013, ApJ, 773, 10
\bibitem[\protect\citeauthoryear{Filippov \& Den}{2000}]{b17}
Filippov B.P., Den O.G., 2000,  Astron. Lett., 26, 322
\bibitem[\protect\citeauthoryear{Filippov \& Den}{2001}]{b18}
Filippov B.P., Den O.G., 2001, J. Geophys. Res., 106, 25177
\bibitem[\protect\citeauthoryear{Filippov, Golub \& Koutchmy}{Filippov et al.}{2009}]{b19}
Filippov, B., Golub, L., Koutchmy, S., 2009, Sol. Phys., 254, 259
\bibitem[\protect\citeauthoryear{Filippov, Koutchmy \& Tavabi}{Filippov et al.}{2013}]{b20}
Filippov, B., Koutchmy, S., Tavabi, E., 2013, Sol. Phys., 286, 143
\bibitem[\protect\citeauthoryear{Heyvaerts, Priest \& Rust}{Heyvaerts et al.}{1977}]{b23}
Heyvaerts, J., Priest, E. R., Rust, D. M., 1977, ApJ, 216, 123
\bibitem[\protect\citeauthoryear{Howard et al.}{2008}]{b24}
Howard R.A. et al., 2008, Space Sci. Rev., 136, 67
\bibitem[\protect\citeauthoryear{Isobe, Tripathi \& Archontis}{Isobe et al.}{2007}]{b25}
Isobe, H., Tripathi, D., Archontis, V., 2007, ApJ, 657, L53
\bibitem[\protect\citeauthoryear{Joshi et al.}{2014}]{b25a}
Joshi N.C., Srivastava A.K., Filippov B., Kayshap P., Uddin W.,
Chandra R., Choudhary D.P., Dwivedi B.N., 2014, ApJ, 787, 11
\bibitem[\protect\citeauthoryear{Kliem \& T$\ddot{o}$r$\ddot{o}$k}{2006}]{b29}
Kliem B.,  T$\ddot{o}$r$\ddot{o}$k T., 2006, Phys. Rev. Lett.,
96(25), 255002
\bibitem[\protect\citeauthoryear{Koutchmy et al.}{1997}]{b31}
Koutchmy, S., Hara, H., Suematsu, Y., Reardon, K., 1997, A\&A,
320, L33
\bibitem[\protect\citeauthoryear{Kurokawa et al.}{1987}]{b32}
Kurokawa, H., Hanaoka, Y., Shibata, K., Uchida, Y., 1987, Sol.
Phys., 108, 251
\bibitem[\protect\citeauthoryear{Lee, Archontis \& Hood}{Lee et al.}{2015}]{b33}
Lee, E.J., Archontis, V., Hood, A.W., 2015, ApJ, 798, L10
\bibitem[\protect\citeauthoryear{Lemen et al.}{2012}]{b34}
Lemen, J. R. et al., 2012, Sol. Phys., 275, 17
\bibitem[\protect\citeauthoryear{Liu et al.}{2009}]{b36}
Liu, W., Berger, T.E., Title, A.M., Tarbell, T.D., 2009, ApJ, 707,
L37
\bibitem[\protect\citeauthoryear{Liu et al.}{2011}]{b37}
Liu, C., Deng, N., Liu, R., Ugarte-Urra, I., Wang, S., Wang, H.,
2011, ApJ, 735, L18
\bibitem[\protect\citeauthoryear{Moore et al.}{2010}]{b40}
Moore, R.L., Cirtain, J W., Sterling, A.C., Falconer, D.A., 2010,
ApJ, 720, 757
\bibitem[\protect\citeauthoryear{Moreno-Insertis \& Galsgaard}{2013}]{b41}
Moreno-Insertis, F., Galsgaard, K., 2013, ApJ, 771, 20
\bibitem[\protect\citeauthoryear{Pariat, Antiochos \& DeVore}{Pariat et al.}{2009}]{b45}
Pariat, E., Antiochos, S.K., DeVore, C.R., 2009, ApJ, 691, 61
\bibitem[\protect\citeauthoryear{Pariat, Antiochos \& DeVore}{Pariat et al.}{2010}]{b46}
Pariat, E., Antiochos, S.K., DeVore, C.R., 2010, 714, 1762
\bibitem[\protect\citeauthoryear{Pariat et al.}{2015}]{b47}
Pariat, E., DeVore, C.R., Dalmasse, K., Antiochos, S.K., Karpen,
J.T., 2015, A\&A, 573, A130
\bibitem[\protect\citeauthoryear{Patsourakos et al.}{2008}]{b47a}
Patsourakos, S., Pariat, E., Vourlidas, A., Antiochos,
S.K.,Wuelser, J.P., 2008, ApJ, 680, L73
\bibitem[\protect\citeauthoryear{Pevtsov, Balasubramaniam \& Roger}{Pevtsov et al.}{2003}]{b48}
Pevtsov, A.A., Balasubramaniam, K.S., Rogers, J.W., 2003, ApJ,
595, 500
\bibitem[\protect\citeauthoryear{Pike \& Mason}{1998}]{b49}
Pike, C.D., Mason, H.E., 1998, Sol. Phys. 182, 133
\bibitem[\protect\citeauthoryear{Rompolt \& Svestka}{1996}]{b50}
Rompolt, B., Svestka, Z., 1996, Adv. Space Res., 17(4/5), 115
\bibitem[\protect\citeauthoryear{Schatten, Wilcox \& Ness}{Schatten et al.}{1969}]{b52}
Schatten, K.H.,Wilcox, J.M., Ness, N.F., 1969, Sol. Phys., 6, 442
\bibitem[\protect\citeauthoryear{Schmieder et al.}{2013}]{b53}
Schmieder, B. et al., 2013, A\&A, 559, A1
\bibitem[\protect\citeauthoryear{Schrijver \& DeRosa}{2003}]{b54}
Schrijver, C.J., DeRosa, M.L., 2003, Sol. Phys., 212, 165
\bibitem[\protect\citeauthoryear{Shen et al.}{2011}]{b55}
Shen, Y., Liu, Y., Su, J., Ibrahim, A., 2011, ApJ, 735, L43
\bibitem[\protect\citeauthoryear{Shen et al.}{2012}]{b56}
Shen, Y., Liu, Y., Su, J., Deng, Y., 2012, ApJ, 745, 164
\bibitem[\protect\citeauthoryear{Shibata}{1998}]{b57}
Shibata, K., 1998, in Proc. Solar Jets and Coronal Plumes, ed.
T.-D. Guyenne (Noodwijk, The Netherlands: ESA Publication
Division), 137
\bibitem[\protect\citeauthoryear{Shibata \& Uchida}{1986}]{b58}
Shibata, K., Uchida, Y., 1986, Sol. Phys., 103, 299
\bibitem[\protect\citeauthoryear{Shibata, Nozawa \& Matsumoto}{Shibata et al.}{1992}]{b59}
Shibata, K., Nozawa, S., Matsumoto, R., 1992, PASJ, 44, 265
\bibitem[\protect\citeauthoryear{Shibata et al.}
{1992}]{b60}
Shibata, K. et al., 1992, PASJ, 44, L173
\bibitem[\protect\citeauthoryear{Shimojo et al.}{1996}]{b61}
Shimojo, M., Hashimoto, S., Shibata, K., Hirayama, T., Hudson,
H.S., Acton, L., 1996, PASJ, 48, 123
\bibitem[\protect\citeauthoryear{Schou et al.}{2012}]{b63}
Schou, J. et al. 2012, Sol. Phys., 275, 229
\bibitem[\protect\citeauthoryear{Steiolfson, Schmahl \& Wu}{Steiolfson et al.}{1979}]{b64}
Steiolfson R.S., Schmahl, E.J., Wu, S.T., 1979, Sol. Phys., 63,
187
\bibitem[\protect\citeauthoryear{Sterling, Harra \& Moore}{Sterling et al.}{2010}]{b66}
Sterling, A.C., Harra, L.K., Moore, R.L., 2010, ApJ, 722, 1644
\bibitem[\protect\citeauthoryear{Suematsu et al.}{1982}]{b67}
Suematsu, Y., Shibata, K., Nishikawa, T., Kitai, R., 1982, Sol.
Phys. 75, 99
\bibitem[\protect\citeauthoryear{Wang \& Liu}{2012}]{b69}
Wang, H., Liu, C., 2012, ApJ, 760, 101
\bibitem[\protect\citeauthoryear{Wuelser et al.}{2004}]{b70}
Wuelser J.-P. et al., 2004, Proc. SPIE, 5171, 111
\bibitem[\protect\citeauthoryear{Xu, Yin \& Ding}{Xu et al.}{1984}]{b71}
Xu, A.-A., Yin, S.-Y., Ding, J.-P., 1984, Acta Astronomica Sinica,
25, 119
\bibitem[\protect\citeauthoryear{Yokoyama \& Shibata}{1996}]{b73}
Yokoyama, T., Shibata, K., 1996, PASJ, 48, 353
\bibitem[\protect\citeauthoryear{Zuccarello et al.}{2014}]{b74}
Zuccarello, F.P., Seaton, D.B., Mierla, M., Poedts, S., Rachmeler,
L.A., Romano, P., Zuccarello, F., 2014, ApJ., 785, 88





\end{thebibliography}
\end{document}